\documentstyle[preprint,aps]{revtex}
\begin{document}
\tighten
\title{On the possibility of $\eta$--mesic nucleus formation\footnote{
Talk given at European Conference on ADVANCES IN NUCLEAR PHYSICS
AND RELATED AREAS, Thessaloniki-Greece 8-12 July 1997}}
\author{ S. A. Sofianos and S. A. Rakityansky}
\address{ Physics Department, University of South Africa,
		P.O.Box 392,Pretoria 0001, South Africa}
\maketitle
\bigskip
Although the $\eta$--meson was discovered 40 years ago, only recently
 particle and nuclear physicists focused their attention on it.
In many respects the $\eta$--meson is similar to the $\pi^0$--meson
despite  it being four times heavier. Both are neutral,
spinless, and have almost the same lifetime, $\sim 10^{-18}$ sec.
The kinship between the two mesons manifests itself very clearly in their
decay modes. They are the only  mesons which have a high probability
of pure radiative decay, i.e.,  their quarks can annihilate into
on-shell photons. The pion almost entirely
decays into the radiative channel $\pi^0\to\gamma+\gamma$
(98.798\,\%).  For the $\eta$
the purely  radiative decay is also the most probable mode \cite{PDG},
$$
\eta\rightarrow\left\{
\begin{array}{ll}
	\gamma+\gamma & (38.8 \%)\\
	\pi^0+\pi^0+\pi^0 & (31.9 \%)\\
	\pi^{+}+\pi^{-}+\pi^0 & (23.6 \%)\\
	\pi^{+}+\pi^{-}+\gamma & (\phantom{0}4.9 \%)\\
	{\rm other  \ decays}&(\phantom{0}0.8 \%)\ \ .\\
\end{array}
\right.
$$
Therefore, when  $\pi^0$ and $\eta$ are viewed as elementary particles,
they look quite similar. However when one considers their interaction
with nucleons, their difference is clearly manifested.
Firstly, one  expects a manifestation of the large $\eta\pi^0$--mass
difference in the meson--nucleon dynamics and,  at low energies,
this is indeed  observed. For example, the $S_{11}$--resonance $N^*(1535)$
is formed in both $\pi N$
and $\eta N$ systems, but at different collision energies,
$$
\begin{array}{lcccrr}
	E^{res}_{\pi N}(S_{11}) &=& 1535\ {\rm MeV}\ - m_{N} -m_{\pi}
	&\approx& 458\ {\rm MeV}&\\
	\phantom{-}&&&&&\\
	E^{res}_{\eta N}(S_{11}) &=& 1535\ {\rm MeV}\ - m_{N} -m_{\eta}
	&\approx& 49\ {\rm MeV}\,.&\\
\end{array}
$$
Note that due to the large mass of the $\eta$--meson (547.45\,MeV),
this resonance is very close to the $\eta N$--threshold. Furthermore
it is  very broad, with
$\Gamma\approx 150\,{\rm MeV}$, covering the whole low energy
region of the $\eta N$ interaction. As a result the interaction of
nucleons with $\eta$--mesons in this region, where the $S$--wave
interaction dominates, is much stronger  than with pions.
Another consequence of the $S_{11}$ dominance is that the
 interaction of the $\eta$--meson with a nucleon can be considered
as a series of formations and decays  of this resonance
as shown in  Fig. 1.
\begin{center}
\unitlength=0.5mm
\begin{picture}(285,70)
\put(135,5){Fig. 1}
\put(30,45){%
\begin{picture}(0,0)%
\multiput(-15,-15)(0.1,0){10}{\line(1,1){12}}
\multiput(43,-15)(-0.1,0){10}{\line(-1,1){12}}
\put(-15,15){\line(1,-1){12}}
\put(43,15){\line(-1,-1){12}}
\multiput(3,-2)(0,0.1){41}{\line(1,0){22}}
\put(0,0){\circle{8}}
\put(28,0){\circle{8}}
\put(-23,13){$\eta$}
\put(-23,-17){$N$}
\put(47,13){$\eta$}
\put(47,-17){$N$}
\put(11,4){$N^*$}
\end{picture}%
}
\put(90,43){+}
\put(130,45){%
\begin{picture}(0,0)%
\multiput(-15,-15)(0.1,0){10}{\line(1,1){12}}
\put(-15,15){\line(1,-1){12}}
\multiput(3,-2)(0,0.1){41}{\line(1,0){22}}
\put(0,0){\circle{8}}
\put(28,0){\circle{8}}
\put(-23,13){$\eta$}
\put(-23,-17){$N$}
\put(40,18){$\eta$}
\put(40,-23){$N$}
\put(11,4){$N^*$}
\multiput(101,-15)(-0.1,0){10}{\line(-1,1){12}}
\put(101,15){\line(-1,-1){12}}
\multiput(61,-2)(0,0.1){41}{\line(1,0){22}}
\put(58,0){\circle{8}}
\put(86,0){\circle{8}}
\put(105,13){$\eta$}
\put(105,-17){$N$}
\put(69,4){$N^*$}
\put(43,4){\oval(30,20)[t]}
\multiput(42.5,-4)(0.1,0){10}{\oval(30,20)[b]}
\end{picture}%
}
\put(250,43){+\quad etc.}
\end{picture}
\end{center}
As with any resonant state, the $N^*(1535)$--resonance
has  branching ratios of the decay modes which do not depend on the
formation channel and after its creation it decays
into $\eta N$ and $\pi N$ channels with equally high probabilities
\cite{PDG}
\begin{equation}
\label{N*decay}
	N^*(1535)\rightarrow\left\{
	\begin{array}{ll}
	N+\eta & (35 - 55\ \%)\\
	N + \pi & (35 - 55\ \%)\\
	{\rm  other\ \ decays}&(\le10\ \%)\ \ \ \ .\\
\end{array}
\right.
\end{equation}
Therefore, the series depicted in Fig. 1, must also include terms describing
real and virtual transitions into the $\pi N$--channel (see Fig 2).
\begin{center}
\unitlength=0.5mm
\begin{picture}(285,70)
\put(135,5){Fig. 2}
\put(30,45){%
\begin{picture}(0,0)%
\multiput(-15,-15)(0.1,0){10}{\line(1,1){12}}
\multiput(43,-15)(-0.1,0){10}{\line(-1,1){12}}
\put(-15,15){\line(1,-1){12}}
\put(43,15){\line(-1,-1){12}}
\multiput(3,-2)(0,0.1){41}{\line(1,0){22}}
\put(0,0){\circle{8}}
\put(28,0){\circle{8}}
\put(-23,13){$\eta$}
\put(-23,-17){$N$}
\put(47,13){$\pi$}
\put(47,-17){$N$}
\put(11,4){$N^*$}
\end{picture}%
}
\put(90,43){+}
\put(130,45){%
\begin{picture}(0,0)%
\multiput(-15,-15)(0.1,0){10}{\line(1,1){12}}
\put(-15,15){\line(1,-1){12}}
\multiput(3,-2)(0,0.1){41}{\line(1,0){22}}
\put(0,0){\circle{8}}
\put(28,0){\circle{8}}
\put(-23,13){$\eta$}
\put(-23,-17){$N$}
\put(40,18){$\pi$}
\put(40,-23){$N$}
\put(11,4){$N^*$}
\multiput(101,-15)(-0.1,0){10}{\line(-1,1){12}}
\put(101,15){\line(-1,-1){12}}
\multiput(61,-2)(0,0.1){41}{\line(1,0){22}}
\put(58,0){\circle{8}}
\put(86,0){\circle{8}}
\put(105,13){$\pi(\eta)$}
\put(105,-17){$N$}
\put(69,4){$N^*$}
\put(43,4){\oval(30,20)[t]}
\multiput(42.5,-4)(0.1,0){10}{\oval(30,20)[b]}
\end{picture}%
}
\put(250,43){+\quad etc.}
\end{picture}
\end{center}
Thus in the energy region covered by the $S_{11}$--resonance, the
$\eta N$ and $\pi N$ interactions should be treated as a coupled channel
problem. When such an analysis was performed, it was found that the
near--threshold $\eta N$ interaction is attractive \cite{bhal}.
This raises the question as to whether this attraction is strong enough
so that  an $\eta$--mesic nucleus can be formed.

Since $\eta$--mesons decay very fast, it is impossible to produce
beams of them and therefore they can only be observed in final states
of certain nuclear reactions with other particles.
This makes investigations of  $\eta$--meson dynamics quite
complicated.  Therefore the possibility of sustaining an $\eta$--meson
inside a nucleus would be an exciting one as
it would expose itself for a relatively
long period in a  series of successive
interactions with nucleons, i.e., inside the nucleus it would
undergo a series of absorptions and emissions through
formations and decays of the $N^*(1535)$--resonance as
depicted in Fig. 3.

\begin{center}
\unitlength=0.5mm
\begin{picture}(215,75)
\put(100,5){Fig. 3}
\put(0,30){%
\begin{picture}(0,0)%
\multiput(30,0)(46,18){3}{%
\begin{picture}(0,0)%
\put(-15,-15){\line(1,1){12}}
\put(0,0){\circle{8}}
\multiput(3,-2)(0,0.1){41}{\line(1,0){22}}
\put(28,0){\circle{8}}
\end{picture}%
}
\put(152,33){%
\begin{picture}(0,0)%
\put(0,0){\line(1,-1){12}}
\put(15,-15){\circle{8}}
\multiput(18,-17)(0,0.1){41}{\line(1,0){22}}
\end{picture}%
}
\multiput(15,-0.5)(0,0.1){10}{\line(1,0){11}}
\multiput(62,-0.5)(0,0.1){10}{\line(1,0){130}}
\multiput(15,17.5)(0,0.1){10}{\line(1,0){57}}
\multiput(108,17.5)(0,0.1){10}{\line(1,0){55}}
\multiput(15,35.5)(0,0.1){10}{\line(1,0){103}}
\multiput(154,35.5)(0,0.1){10}{\line(1,0){38}}
\multiput(6,-2)(0,18){3}{$N$}
\put(7,-17){$\eta$}
\put(195,-2){$N$}
\put(195,16){$N^*$}
\put(195,34){$N$}
\end{picture}%
}
\end{picture}
\end{center}

The lifetime of such an $\eta$--mesic nucleus would not
be limited by the lifetime of the \mbox{$\eta$--meson} itself because after
each creation of the $S_{11}$--resonance the $\eta$--meson is
generated anew. However, such an $\eta$--nucleus state
can not be stable, since eventually the
$N^*(1535)$--resonance will produce a pion instead of $\eta$
as their creation probabilities are, according to (\ref{N*decay}),
equally high.
Of course such a pion can generate an $N^*(1535)$--resonance again which in
turn may revive the $\eta$ but such a possibility is rather low since
the pion acquires, through the decay of the resonance, a kinetic energy
of $\sim 400$\,MeV and  can thus easily escape.
It is therefore clear that if an $\eta$--meson is bound inside a
nucleus, it can only be in a  quasi--bound state
 with nonzero width.

First estimation, obtained in the framework of the optical potential
theory, put a lower bound on the number of nucleons $ A$ which is
necessary to bind the $\eta$--meson, namely, $ A\ge 12$ \cite{haider}.
Thereafter other theoretical investigations were devoted to this
problem. All of them predicted $\eta$--nucleus bound states obeying this
constraint. However, the search for narrow $\eta$-nuclear bound states
in an experiment with lithium, carbon, oxygen, and aluminum by
Chrien et al. \cite{chrien} produced negative results.

The conclusion of this experimental work, however, did not
discourage theoreticians in examining the possibility of an
$\eta$--nucleus formation. The  relatively large scattering lengths
obtained for $\eta {}^3$He and $\eta {}^4$He
systems  using a zero--range $\eta N$--interaction \cite{Wilk}
cast doubt on the $ A\ge 12$ constraint.
Speculations of this kind are based on the argument
that in the vicinity of the origin of the complex momentum plane
the amplitude $f$ can be replaced by the scattering length $a$
and therefore the $S$--matrix in this area can be written as
$$
	S=1+2ikf\approx1+2ika\approx \frac{1+ika}{1-ika}\ .
$$
This expression  is valid only for small $k$ and  can have a pole
in this region only if $a$ is large. If $a$ is negative the  pole
would be on the positive imaginary axis (bound state).
Thus, a large negative scattering length indicates that
a weakly bound state exists.
Decreasing the interaction strength transforms the
bound state into a resonance, and vice versa, implying a
change of the scattering length from $-\infty$ to $+\infty$.

Such simple reasoning, however, is valid only when the interaction is
described by a real potential. In the case of an $\eta$--nucleus
system the inelastic $\eta A\to \pi A$  channel is always open
giving rise to a significant imaginary part in the $\eta$--nucleus potential.
The resonance and quasi--bound state poles of the $S$--matrix generated by a
complex potential have quite different distribution in the complex
$k$--plane. In Ref. \cite{cassing} it was shown  that starting from
a purely real potential and introducing an imaginary part
which is gradually increased, results in $S$--matrix pole-behaviour
shown in Fig. 4.
\begin{center}
\unitlength=0.5mm
\begin{picture}(300,100)
\put(50,5){Fig. 4}
\put(60,60){%
\begin{picture}(0,0)%
\put(-40,0){\line(1,0){90}}
\put(0,-40){\line(0,1){80}}
\put(0,0){\line(-1,1){40}}
\put(0,15){\circle*{4}}
\put(0,25){\circle*{4}}
\put(15,-10){\circle*{4}}
\put(25,-15){\circle*{4}}
\put(-15,-10){\circle*{4}}
\put(-25,-15){\circle*{4}}
\put(-25,-15){\vector(-1,2){20}}
\put(-15,-10){\vector(-1,2){10}}
\put(15,-10){\vector(0,-1){15}}
\put(25,-15){\vector(0,-1){20}}
\put(0,15){\vector(-2,1){15}}
\put(0,25){\vector(-2,1){25}}
\put(3,35){${\rm Im\,}k$}
\put(40,-7){${\rm Re\,}k$}
\put(-40,-26){\it resonances}
\put(30,-26){\it resonances}
\put(5,21){\it bound}
\put(5,13){\it states}
\end{picture}%
}
\put(160,0){%
\begin{picture}(0,0)
\put(60,5){Fig. 5}
\put(70,60){%
\begin{picture}(0,0)%
\put(-50,20){\circle{5}}
\put(-50,-20){\circle{5}}
\put(-30,0){\circle{5}}
\put(10,20){\circle{5}}
\put(10,-20){\circle{5}}
\put(40,20){\circle{5}}
\put(40,-20){\circle{5}}
\put(-50,-17.5){\vector(0,1){35}}
\put(-32.5,0){\vector(-1,0){17.5}}
\put(10,-17.5){\vector(0,1){35}}
\put(40,-17.5){\vector(0,1){35}}
\put(40,0){\vector(-1,0){30}}
\put(-54,25){$N_1$}
\put(-54,-31){$N_2$}
\put(-25,-3){$N_3$}
\put(6,25){$N_1$}
\put(6,-31){$N_2$}
\put(36,-31){$N_4$}
\put(36,25){$N_3$}
\put(-60,-2){$\vec x_1$}
\put(-45,3){$\vec x_2$}
\put(0,-2){$\vec x_1$}
\put(20,3){$\vec x_2$}
\put(43,-2){$\vec x_3$}
\end{picture}%
}
\end{picture}%
}
\end{picture}
\end{center}
Therefore, in the case of a
complex potential, both resonance and quasi--bound state poles are
situated in the second quadrant of the complex momentum plane,
under and above its diagonal respectively.
The diagonal separates them because the energy $E_0=k_0^2/2\mu$
corresponding to a pole at $k=k_0$,
$$
	E_0=\frac{1}{2\mu}\left[({\rm Re\,}k_0)^2-({\rm Im\,}k_0)^2+
	2i({\rm Re\,}k_0)({\rm Im\,}k_0)\right]\ ,
$$
has a positive (negative) real part when $k_0$ is under (above) it.
Thus, the transition from resonances to quasi--bound states is a crossing
of the diagonal. Since this can take place rather far from the point
$k=0$, we should not expect, in contrast to the  real
potential case, to have a large scattering length even if the
binding energy, $|{\rm Re\,}E_0|$, is small. Moreover,
crossing the diagonal is not associated with dramatic changes of $a$.
 In short, scattering length calculations cannot
provide a definite answer and a more rigorous approach must be employed.
The most adequate way to solve this problem is to locate the poles
of the $S$-matrix in the second quadrant of the $k$-plane. In Refs.
\cite{ours,Raki4} we  developed a microscopic method
 that enabled us to calculate the elastic scattering amplitude for
any complex value of $k$ and thereby to locate its poles. The influence
of  inelastic channels is taken into account via a complex $\eta$N
potential. In what follows this method is described in somewhat
more detail.


Consider the scattering of an $\eta$-meson from a nucleus consisting
of  $A$ nucleons. The Hamiltonian of the system is given by
\begin{equation}
	  H = H_0 + V_{\eta A} + H_A
\label{H1}
\end{equation}
where
$H_0$ is the  free Hamiltonian corresponding to the  $\eta$--nucleus
relative motion,
$ V_{\eta A} = V_1 + V_2 +\cdots+ V_A $ is  the sum of  $\eta N$ potentials,
    $V_i\equiv V_{\eta N}(|\vec R-\vec r_i|)$,
where $\vec R$  and $\vec r_i$ are the coordinates of the $\eta$
 and the $i$-th nucleon with respect to the $c.m.$
of the nucleus, and $H_A$ is  the total Hamiltonian of the nucleus,
\begin{equation}
	H_A=\frac {\hbar}{i}\sum_{i=1}^A\nabla_{\vec r_i}+\sum_{i \ne j}
		V_{NN}(|\vec r_i-\vec r_j|)\,.
\label{HA}
\end{equation}
The elastic scattering amplitude $f(\vec k',\vec k;z)$
describing the transition
from the initial,  $|\vec{k},\psi_0\rangle$,  to the final,
 $|\vec{k}',\psi_0\rangle$, asymptotic state where $|\psi_0\rangle$ is
the nuclear  ground state and $\vec k$ the $\eta$-nucleus
relative momentum, can be expressed in terms of the T--matrix elements
\begin{equation}
       f(\vec{k}',\vec{k}; z)=-\frac{\mu}{2\pi}\
      <\vec{k}',\psi_0|T(z)|\vec{k},\psi_0>\ .
\label{ampli}
\end{equation}
The operator  $T$ is related to the Green function
$G_A(z)=(z-H_0-H_A)^{-1}$ by
\begin{equation}
	T(z)=V+VG_A(z)T(z)\ .
\label{tmat}
\end{equation}
The task of solving Eq. (\ref{tmat}) is a formidable one and thus
one must resort to approximations. One such approximation
is the so--called Finite-Rank Approximation (FRA) of the Hamiltonian. It
has been proposed in Refs. \cite{Bel1,Bel2} as an alternative to the
multiple scattering and optical potential theories. In this method the
auxiliary  operator
\begin{equation}
       T^0(z)=V+VG_0(z)T^0(z)\,,
\label{tmat0}
\end{equation}
where $ G_0(z)=(z-H_0)^{-1}$ is the free Green function, is introduced.
Using the identity
$
	A^{-1}-B^{-1}=B^{-1}(B-A)A^{-1}
$
with    $A=z-H_0-H_A$ and $B=z-H_0$, one gets the resolvent equation
\begin{equation}
      G_A(z)=G_0(z)+G_0(z)H_A G_A(z)\,,
\label{GA}
\end{equation}
and thus
\begin{equation}
       T(z)=T^0(z)+T^0(z) G_0(z)H_A G_A(z)T(z)\,.
\label{THA}
\end{equation}
The latter equation has the advantage that the spectral
decomposition for the Hamiltonian,
\begin{equation}
    H_A=\sum_n {\cal E}_n |\psi_n><\psi_n| + \int dE\,E|\psi_E><\psi_E|\,,
\label{Hspace}
\end{equation}
can be employed to bring Eq. (\ref{tmat})  into a manageable form.
The FRA method  is based on the approximation
\begin{equation}
     H_A\approx {\cal E}_0|\psi_0><\psi_0|
\label{Happrox},
\end{equation}
which  means that during the scattering of the $\eta$-meson,
the nucleus remains  in its ground state $|\psi_0>$.
Such an approximation is widely used in multiple scattering and
optical potential theories where is known as the coherent approximation.
Using  (\ref{Happrox})  we get
\begin{equation}
      T(z)=T^0(z)+{\cal E}_0T^0(z)|\psi_0> G_0(z)G_0(z-{\cal E}_0)
		<\psi_0|T(z)\,.
\label{tmata}
\end{equation}
%
The matrix elements $T({\vec k}',{\vec k};z)
\equiv <\vec{k}',\psi_0|T(z)|\vec{k},\psi_0>$ are thus given by
\begin{eqnarray}
\nonumber
   T(\vec{k}',\vec{k};z) &=& <\vec{k}',\psi_0|T^0(z)|\vec{k},\psi_0>\\
  &+&
       {\cal E}_0\int \frac{d\vec{k}''}{(2\pi)^3}
	\frac{ <\vec{k}',\psi_0|T^0(z)|\vec{k}'',\psi_0>}{
       (z-{k''}^2/2\mu)(z-
	{\cal E}_0-{k''}^2/2\mu)}\,T(\vec{k}'',\vec{k};z)\,.
\label{tm3}
\end{eqnarray}
The auxiliary operator $T^0$ describes the scattering of the $\eta$-meson
from nucleons fixed in their space position within the nucleus. This is
clear since Eq. (\ref{tmat0}) does not contain any operator acting on
the internal nuclear Jacobi coordinates denoted by
 $\{\vec{r}\}\equiv\{\vec x_1,\vec x_2,\cdots,\vec x_{A-1}\}$.
Therefore all operators in Eq. (\ref{tmat0}) are diagonal
in the configuration subspace \{$\vec{r}$\} and thus
\begin{equation}
	T^0(\vec{k}',\vec{k};\vec{r};z)=V(\vec{k}',\vec{k};\vec{r})
	+ \int \,\frac{d\vec{k}''}{(2\pi)^3}
	\frac{V(\vec{k}',\vec{k}'';\vec{r})}
	{z-k''^2/2\mu}\,T^0(\vec{k}'';\vec{k};\vec{r};z)
\label{t0m}
\end{equation}
where
$$
   <\vec{k}',\vec{r}\,'|T^0(z)|\vec{k},\vec{r}>  =
     \delta(\vec{r}\,'-\vec{r})\, T^0(\vec{k}',\vec{k};\vec{r};z)\,,\quad
<\vec{k}',\vec{r}\,'|V|\vec{k},\vec{r}>  =  \delta(\vec{r}\,
    '-\vec{r})\,V(\vec{k}',\vec{k}; \vec{r})\,.
$$
It is clear that $T^0(\vec{k}',\vec{k};\vec{r};z)$ depends parametrically
on  $\{\vec{r}\}$.
Therefore the matrix elements $<\vec{k}^\prime, \psi_0 | T^0(z)|\vec{k},
\psi_0>$ can be obtained by integrating over the Jacobi coordinates
\begin{equation}
	<\vec{k}^\prime, \psi_0 | T^0(z) | \vec{k}, \psi _0 >
	= \int d\vec{r} |
	\psi_0(\vec{r})|^2 T^0(\vec{k}^\prime,\vec{k};\vec{r};z)\,.
\label{aver}
\end{equation}
Thus  the solution of the scattering problem can be obtained by solving
first Eq. (\ref{t0m}), averaging  as in Eq. (\ref{aver}), and
finally calculating $T$ from Eq. (\ref{tm3}). We must  emphasize that the
above scheme, is not the same as the first order optical potential
approach used in the traditional
pion-nucleus multiple scattering theory \cite{Land}.
Indeed, the latter is based on three approximations; i) the Impulse
Approximation; ii) the omission of higher order rescattering terms in
constructing the optical potential, and iii)  the coherent approximation.
In contrast, in the scheme considered here, the Impulse Approximation to
obtain the $\eta$N amplitude in nuclear media is not needed
and no rescattering terms are omitted.

The parameter $z$ in the above equations corresponds to the total
 $\eta$-nucleus  energy, $ z = E - |{\cal E}_0| + i0$,
where $E$ is the energy associated with the $\eta$-nucleus relative-motion.
On the energy shell we have $E = k^2/2\mu$.  Therefore, even the  auxiliary
$T^0$-matrix differs from the conventional
fixed-scatterer amplitude in that it is always taken off the
energy shell. In the case of scattering length calculations, we have
$E=0$ and thus $z = -|{\cal E}_0|$. This makes Eqs.
(\ref{tm3}) and (\ref{t0m}) nonsingular and easy to handle.

For  practical calculations we rewrite Eq. (\ref{t0m}) using the
Faddeev-type decomposition
\begin{eqnarray}
\nonumber
	T^0(\vec{k}',\vec{k};\vec{r};z) &=& \sum^A_{i=1} T^0_i
	(\vec{k}',\vec{k};\vec{r};z),\\
	T^0_i (\vec{k}',\vec{k};\vec{r};z) &=&
	t_i(\vec{k}',\vec{k};\vec{r};z) \
	+ \int \frac {d\vec{k}''}{(2\pi)^3}
	\frac {t_i(\vec{k}',\vec{k}'';\vec{r};z)} {z - {k''}^2/2\mu}
	\sum_{j\neq i} T^0_j(\vec{k}'',\vec{k};\vec{r};z)\ ,
\label{t0i}
\end{eqnarray}
where  $t_i$ is the t-matrix for the $\eta$-meson
scatterred by the nucleon  $i$ and is expressed in terms of
the two-body $t_{\eta N}$-matrix via
\begin{equation}
	t_i(\vec k',\vec k;\vec r;z)=t_{\eta N}(\vec k',\vec k;z)
	 \exp\left[{{\displaystyle i(\vec k-\vec k')\cdot\vec r_i}}\right]\,.
\label{telem}
\end{equation}
Expanding the $|\vec{k}, \vec{r} >$ basis in partial waves
and using the fact that, at the low energies considered
here, the $\eta$N interactions is dominated by the S$_{11}$--resonance,
we may retain the S-wave only. The total orbital momentum
is zero and since the $\eta$-meson is a spinless particle
the nuclear spin can be ignored. Therefore, when Eq. (\ref{t0i})
is projected on the  S-wave basis $|k, r >$, it reduces to
\begin{equation}
 T^0_i(k', k;r; z) = t_i(k', k;r;z) + \frac {1}{2\pi^2} \int^{\infty}_0\,dk''
   \frac {k''^2\,t_i(k',k'';r;z)}{z-k''^2/2\mu}\,\sum_{j\neq i}\ T^0_j
	(k'',k;r,;z)
\label{A4}
\end{equation}
where
$$
	< k', r' | T^0_i (z) |k, r > =
		\frac {\delta (r'-r)}{4\pi r^2}\ T^0_i (k',k;r; z)
$$
and similarly for $< k',r' | t_i (z) | k,r >$.

The above formulae are given for the general case of $A$ nucleons.
In what follows we restrict ourselves to $A=$2, 3, and 4. The relevant
Jacobi vectors are shown in Fig. 5.  According to Eq. (\ref{telem}), $t_i$
depends on the space configuration of the nucleons because $k$ and
$k'$ are the $\eta$-meson momenta with respect to the nuclear centre
of mass  while the
nucleon $i$ is shifted from it  by the vector $
	\vec{r}_i = a_i\vec{x_1} + b_i\vec{x_2} + c_i\vec{x_3}\,,
$ where
$a_1 = \frac {1}{2}, a_2 = - \frac {1}{2}, b_1 = b_2 = c_1 = c_2 = 0$ for
the deuteron case; $a_1 = \frac {1}{2},\, a_2 = - \frac {1}{2},\, a_3 =
0,\, b_1 = b_2 = \frac {1}{3},\, b_3 = - \frac {2}{3},\, c_1 = c_2 =
c_3 = 0$ for the three-nucleon case, and $a_1 = \frac {1}{2},\, a_2 =
- \frac {1}{2},\, a_3 = a_4 = 0,\, b_1 = b_2 = \frac {1}{2},\,
b_3 = b_4 = - \frac {1}{2},\, c_1 = c_2 = 0, c_3 = \frac {1}{2},\,
 c_4 = - \frac {1}{2}$ for the four-nucleon case.

The $S$-wave projection of Eq. (\ref{telem}) gives
\begin{eqnarray*}
	\langle k',r' &|& t_i(z)|k,r \rangle= \int \, \frac
		{d\vec{k}_i'd\vec{k}_i}{(2\pi
		)^6} d\vec{r}\,''d\vec{r}\,'''\langle k',r'
		| \vec{k}_i',\vec{r}\,''\rangle
		\langle\vec{k}_i',\vec{r}\,''|
		t_i(z)| \vec{k}_i,\vec{r}\,'''\rangle
		\langle\vec{k}_i,\vec{r}\,'''|k,r\rangle \\
		& =&  \frac{\delta(r'-r)}{4\pi r^2}\,j_0(a_ik'x_1)
		j_0(b_ik'x_2)  j_0(c_ik'x_3)
		 \,t_{\eta N}(k',k;z)\,j_0(a_ikx_1)j_0(b_ikx_2)j_0(c_ikx_3)\,,
\end{eqnarray*}
where $j_0$ is the spherical Bessel function.
The $\eta N$ interaction can be described by the $t$-matrix
\begin{equation}
	   t_{\eta N}(k',k;z) = \frac {\lambda}{(k'^2+
	   \alpha^2)(z - E_0 + i\Gamma/2)(k^2+\alpha^2)}\,.
	   \label{tnN}
\end{equation}
This ansatz is motivated by the $S_{11}$--resonance dominance. The
vertex function
for $\eta$N\,$\leftrightarrow$\,N$^*$ is chosen as $1/(k^2+\alpha^2)$
which in configuration space has a Yukawa-type behaviour.
The propagator is taken to be of a simple Breit-Wigner form.
With such a choice the $t_i$ has the following separable form
\begin{equation}
\label{set1}
	t_i(k',k,r;z)= H_i(k';r)\,\tau (z) \,H_i(k,r)\\
\end{equation}
where
\begin{equation}
\label{set2}
  \tau (z)  =  \frac{\lambda}{z-E_0+i\Gamma /2}\,,\qquad
	H_i(k,r)=\frac{j_0(a_ikx_1)j_0(b_ikx_2)j_0(c_ikx_3)}
		{(k^2+\alpha^2)}\,.
\end{equation}
Therefore,
\begin{equation}
	T^0(k',k,r;z) = \sum_{i,j=1}^A\,H_i(k';r)\,
		\Lambda_{ij}(z)\,H_j(k,r)
\end{equation}
where
\begin{equation}
	(\Lambda^{-1})_{i,j} = \frac{\delta_{ij}}{\tau(z)}
		-(1-\delta_{ij})\Gamma_{ij}(r,z)\,,\qquad
		\Gamma_{ij}(r,z)=\frac{1}{2\pi^2}\int_0^\infty\,
		dk\frac{k^2}{z-k^2/2\mu} H_i(k;r)H_j(k,r)\,.
\end{equation}
Formally, the  the last integral involves products of
six Bessel functions. However, several of the coefficients
 $a_i$, $b_i$, and $c_i$ are  always zero and therefore
only products of at most  four Bessel functions can  appear
in the expression for  $\Gamma_{ij}(r,z)$ with the required integrals
having the general form
\begin{equation}
		\gamma(p,u,v,w)=\int_0^\infty dk\frac{k^2j_0(ku)
		j_0(kv)[j_0(kw)]^2}{(k^2+\alpha^2)^2(k^2-p^2-i0)}\,.
\end{equation}
To calculate the latter integral we introduce the auxiliary one
\begin{equation}
	\hat{ \gamma}(p,u,v,w,\delta )=\int_0^\infty dk \frac
	{k^2j_0(ku)j_0(kv)[\sin(kw)]^2} {(k^2+\alpha^2)^2(k^2-p^2-i0)(k^2
	+\delta^2)w^2}
\end{equation}
and thereafter evaluate the limit
\begin{equation}
	\gamma (p,u,v,w)= \lim_{\delta \rightarrow 0} \hat{\gamma }
		(p,u,v,w,\delta )\,.
\end{equation}
The result thus obtained is
\begin{eqnarray}
\nonumber
	\gamma(p,u,v,w)&=&\frac{1}{16uvw^2} [g(u+v+2w)-2g(u+v)+g(u+v-2w)\\
	& &-g(u-v+2w) + 2g(u-v)-g(u-v-2w)]\,,
\end{eqnarray}
where
\begin{eqnarray}
\nonumber
	g(s) &=& \frac {i\pi }{ (p^2 + \alpha^2 )^2}
	\bigg \{{\rm sign} ({\rm Im}\ p) \,\frac{1}{p^3}
	\exp{ [ip|s| {\rm sign} ({\rm Im}\ p)]}\\
	& & -\frac{i\exp{ (-\alpha |s|)}}{2 \alpha^5}[2\alpha^2
	+(3+\alpha |s|)(p^2+\alpha^2)]-
	\frac {i|s|(p^2+\alpha^2)^2}{\alpha^4 p^2}\bigg\}
\end{eqnarray}
with
\begin{equation}
	{\rm sign}(\alpha)=\cases {+1, & for $\alpha \geq 0$ \cr
	-1,& for $\alpha <$ 0\ . \cr}
\end{equation}

To obtain the necessary nuclear wave  functions $\psi_0$ we employed
the  Malfliet--Tjon $NN$--potential \cite{mt} and the
integro--differential equation approach  (IDEA) \cite{idea1,idea2} which,
for $S$--wave projected potentials, is equivalent to the exact Faddeev
equations.

Using the above  formalism, the position and movement of poles of
the $\eta$--meson--light nuclei
\mbox{ ($^2$H, $^3$H, $^3$He, and $^4$He)}
elastic scattering amplitude in the complex $k$-plane
are  studied.
The two-body t-matrix is assumed to be of the form (\ref{tnN}) with
$E_0 = 1535$ MeV\,$- \,(m_N + m_{\eta})$
and  $\Gamma = 150$ MeV \cite{PDG}.
The  range parameter used,  $\alpha =2.357$~fm$^{-1}$,
was obtained  in a two--channel fit to the $\pi N\rightarrow\pi N$ and
$\pi N\rightarrow\eta N$ experimental data \cite{bhal}.
The parameter $\lambda$ is chosen to provide the correct zero-energy
on-shell
limit, i.e., to reproduce the $\eta N$ scattering length $a_{\eta N}$,
$$
 t_{\eta  N}(0,0,0) = - \frac {2\pi}{\mu_{\eta  N}}a_{\eta  N}\,.
$$
The scattering length $a_{\eta N}$, however,  is not  accurately known.
Different analyses \cite{batinic} provided values for the real part in the
range Re\,$a_{\eta N}\in [0.27,0.98]\, {\rm fm}$ and for the imaginary
part
${\rm Im\,}a_{\eta N}\in [0.19,0.37]\,{\rm fm}$.
Therefore in a search for
 bound states one must  use values of $a_{\eta N}$ within these ranges.
To achieve this we used $a_{\eta N}=(\zeta 0.55+i0.30)$ fm
and vary $\zeta$ until a bound state appears.

The poles found with $a_{\eta N}=(0.55+i0.30)$ fm are
shown in Fig. 6. The corresponding energies and widths are given in Table 1.
When ${\rm Re\,}a_{\eta N}$ increases, all the poles
move up and to the right, and when a resonance pole crosses the
diagonal it becomes a quasi--bound pole. The minimal values of
${\rm Re\,}a_{\eta N}$ which generate `zero--binding'
(the poles just on the diagonal) are given in  Table 2.

All these values are within the uncertainty interval
${\rm Re\,}a_{\eta N}\in [0.27,0.98]$ fm. Thus even the possibility
of an $\eta$d binding cannot be at present excluded. Most recent
estimates of ${\rm Re\,}a_{\eta N}$ \cite{newa} are concentrated
around the value ${\rm Re\,}a_{\eta N}\approx 0.7$ fm, which enhances
our belief that at least the  $\alpha$--particle can entrap  an
$\eta$--meson.

\begin{center}
\unitlength=0.5mm
\begin{picture}(290,85)
\put(110,30){%
\begin{picture}(0,0)%
\put(-50,-30){Fig. 6}
\put(0,0){\line(-1,0){100}}
\put(0,0){\line(0,1){40}}
\multiput(0,-2)(-10,0){11}{\line(0,1){2}}
\put(0,-4){\line(0,1){2}}
\put(-50,-4){\line(0,1){2}}
\put(-100,-4){\line(0,1){2}}
\multiput(2,0)(0,10){5}{\line(-1,0){2}}
\put(-20,45){${\rm Im\,}p\, ({\rm fm}^{-1})$}
\put(-105,-15){${\rm Re\,}p\, ({\rm fm}^{-1})$}
\put(-58,-12){-0.5}
\put(0,0){\line(-1,1){40}}
\put(4,8){0.1}
\put(4,18){0.2}
\put(4,28){0.3}
\put(4,-2){0}
\put(-1,-12){0}
\put(-90.259,35.870){\circle{3.00}}
\put(-56.045,23.859){\circle{3.00}}
\put(-54.692,24.478){\circle{3.00}}
\put(-16.504,27.876){\circle*{3.00}}
\put(-90,40){${}^2{\rm H}$}
\put(-70,20){${}^3{\rm H}$}
\put(-50,25){${}^3{\rm He}$}
\put(-15,17){${}^4{\rm He}$}
\put(75,15){%
\begin{tabular}{|c|c|c|}
\hline
system & $E$\ (MeV) & $\Gamma^{\mathstrut}_{\mathstrut}$\ (MeV)\\
\hline
$\eta\ {}^2{\rm H}$ & 31.46 & 59.38 \\
\hline
$\eta\ {}^3{\rm H}$ & 10.91 & 22.68 \\
\hline
$\eta\ {}^3{\rm He}$& 10.14 & 22.70 \\
\hline
$\eta\ {}^4{\rm He}$& -2.05 & 7.48 \\
\hline
\end{tabular}%
}
\end{picture}%
}
\put(225,0){Table 1}
\end{picture}
\end{center}
For each of the four nuclei considered, the scattering lengths
were calculated with eight values of the strength parameter
$\lambda$ corresponding to  ${\rm Re\,}a_{\eta N}$:
$\{(0.2+0.1n)\ {\rm fm}; n=1,8\}$, which extends over the uncertainty
interval. The ${\rm Im\,}a_{\eta N}$ was fixed to the value 0.3 fm.
An increase of ${\rm Re\,}a_{\eta N}$ moves the points
along the trajectories, shown Figs. 7 and 8, anti-clockwise. When
${\rm Re\,}a_{\eta N}$ exceeds the critical values given in Table 2,
 the $\eta N$
interaction becomes strong enough to generate a quasi--bound state.
The corresponding $\eta$--nucleus scattering lengths
 are  shown by filled circles (the trajectories for $^3$He and $^3$H
are practically the same).

\begin{center}
\unitlength=0.45mm
\begin{picture}(280,110)
\put(10,25){%
\begin{picture}(0,0)%
\put(-40,0){\line(1,0){70}}
\put(0,0){\line(0,1){70}}
\multiput(-40,-2)(10,0){8}{\line(0,1){2}}
\multiput(-2,10)(0,10){7}{\line(1,0){2}}
\put(40,-10){\llap{${\rm Re}\,a_{\eta A}$}}
\put(-2,75){\llap{${\rm Im}\,a_{\eta A}$}}
\put(-44,-10){-4}
\put(-34,-10){-3}
\put(-24,-10){-2}
\put(-14,-10){-1}
\put(-2,-10){0}
\put(8,-10){1}
\put(17,12){${}^2{\rm H}$}
\put(-42,35){${}^4{\rm He}$}
\put(5.60,11.91){\circle{3.00}}
\put(8.73,15.07){\circle{3.00}}
\put(12.10,19.66){\circle{3.00}}
\put(15.22,26.43){\circle{3.00}}
\put(16.64,36.21){\circle{3.00}}
\put(13.40,48.53){\circle{3.00}}
\put(3.01,59.18){\circle{3.00}}
\put(-11.25,62.50){\circle*{3.00}}
\put(-11.91,31.00){\circle{3.00}}
\put(-26.13,32.69){\circle{3.00}}
\put(-35.73,23.09){\circle*{3.00}}
\put(-35.94,13.60){\circle*{3.00}}
\put(-32.98,8.45){\circle*{3.00}}
\put(-30.05,6.33){\circle*{3.00}}
\put(-28.42,5.50){\circle*{3.00}}
\put(-26.41,4.27){\circle*{3.00}}
\put(-10,-25){Fig. 7}
\end{picture}%
}
\put(130,25){%
\begin{picture}(0,0)%
\put(-40,0){\line(1,0){70}}
\put(0,0){\line(0,1){70}}
\multiput(-40,-2)(10,0){8}{\line(0,1){2}}
\multiput(-2,10)(0,10){7}{\line(1,0){2}}
\put(40,-10){\llap{${\rm Re}\,a_{\eta A}$}}
\put(-2,75){\llap{${\rm Im}\,a_{\eta A}$}}
\put(-44,-10){-4}
\put(-34,-10){-3}
\put(-24,-10){-2}
\put(-14,-10){-1}
\put(-2,-10){0}
\put(8,-10){1}
\put(-35,47){${}^3{\rm H}$}
\put(-25,33){${}^3{\rm He}$}
\put(2.21,23.48){\circle{3.00}}
\put(0.82,33.01){\circle{3.00}}
\put(-7.38,43.13){\circle{3.00}}
\put(-22.07,46.1){\circle{3.00}}
\put(-33.65,39.41){\circle{3.00}}
\put(-37.84,30.71){\circle*{3.00}}
\put(-37.99,24.28){\circle*{3.00}}
\put(-36.83,20.19){\circle*{3.00}}
\put(2.03,23.43){\circle{3.00}}
\put(0.60,32.80){\circle{3.00}}
\put(-7.42,42.63){\circle{3.00}}
\put(-21.57,45.60){\circle{3.00}}
\put(-37.26,31.17){\circle*{3.00}}
\put(-37.69,24.92){\circle*{3.00}}
\put(-36.77,20.88){\circle*{3.00}}
\put(-10,-25){Fig. 8}
\end{picture}%
}
\put(200,57){%
\begin{tabular}{|c|c|}
\hline
system & min$\{{\rm Re\,}a_{\eta N}\}^{\mathstrut}_{\mathstrut}$\ (fm)\\
\hline
$\eta {}^2{\rm H}^{\mathstrut}_{\mathstrut}$ & 0.91 \\
\hline
$\eta {}^3{\rm H}$ & 0.75 \\
\hline
$\eta {}^3{\rm He}$ & 0.73 \\
\hline
$\eta {}^4{\rm He}$ & 0.47 \\
\hline
\end{tabular}%
}
\put(243,0){Table 2}
\end{picture}
\end{center}
Finally, we would like to emphasize that the  spectral properties of
Hermitian and non-Hermitian Hamiltonians are quite different. Locating
quasi--bound states is a delicate problem which can be treated only
by rigorous methods. As we have shown in Ref. \cite{Raki4} the
$\eta$A scattering length tells us nothing about the existence or
not of an $\eta$A quasi--bound state. This is clearly seen on Figs. 7
and 8 where the trajectories go smoothly from open to filled circles
without any drastic changes or extreme values.

In summary, it is  shown that within  the existing uncertainties
of the elementary  $\eta$N iteraction all light nuclei considered
can support a quasi--bound state which can result in
an $\eta$-mesic nucleus  which is analogous to hypernuclei. Due to
the specific quantum  numbers of the $\eta$--meson (I=0, S=0) such states,
if they do exist, can be used to access new nuclear states
inaccessible by other mesons such as pions and kaons.
Furthermore it can be used to elucidate the role played by the
$\eta$ meson in Charge Symmetry Breaking reactions  and in the
violation of the Okubo-Zweig-Iizuka (OZI) rule.
\vspace*{-20cm}

\end{document}